\documentstyle[sprocl]{article}
\def\Journal#1#2#3#4{{#1} {\bf #2}, #3 (#4)}

\def\APJ{{\em Ap. J.}}

\def\MNRAS{{\em Mon. Not. Royal. Astron. Soc.}}

\def\be{\begin{equation}}
\def\ee{\end{equation}}
\def\bea{\begin{eqnarray}}
\def\eea{\end{eqnarray}}

\newcount\eqnumber
\def\neweq{{\the\eqnumber}\global\advance\eqnumber by 1}
\def\eqnam#1#2{\xdef#1{\the\eqnumber}}

\def\lasteq{\advance\eqnumber by -1 {\the\eqnumber}\advance
     \eqnumber by 1}
\begin{document}
\title{Limits on  Expanding Relativistic Shells from Gamma-Ray Burst
Temporal Structure}
\author{ E. E. FENIMORE}
\address{Los Alamos National Laboratory, MS D436,\\ Los Alamos, NM 87544, USA}
\maketitle\abstracts{
We calculate the expected envelope of emission for relativistic shells
under the assumption of local spherical symmetry.  
Gamma-Ray Burst
envelopes rarely conform to the expected shape, which has 
a fast rise and a smooth, slower decay. Furthermore,
the duration of the decay phase
is related to the time the shell expands before converting its energy
to gamma rays.  From this, one can estimate the energy required for the
shell to sweep up the ISM. The energy greatly exceeds $10^{53}$ erg
unless the bulk Lorentz factor is less than 75.  This puts extreme
limits on the ``external'' shock models. 
However, the alternative, ``internal'' shocks from a central engine,  has
one large
problem: the entire long complex time history lasting hundreds of
seconds
must be
postulated at the central site.}
 
The temporal structure of long complex Gamma-Ray Bursts (GRBs) presents a
myriad of problems for models that involve a single central release of
energy, as in many cosmological scenarios.  Bursts with 50 peaks within
100 seconds are not uncommon, and there is the recent
report~\cite{repeat}
of a burst
which might have lasted from $10^3$ to $10^5$ seconds.
In Fenimore, Madras, \& Nayakshin~\cite{fmn} (hereafter FMN), we 
used kinematic limits
and the observed temporal structure of GRBs to estimate the 
characteristics of the gamma-ray producing regions.  
The bulk Lorentz factor
of the shell, $\Gamma$, must be $10^2$ to
$10^3$ in order to avoid photon-photon attenuation~\cite{baring,nf}.
 Since the emitting surface is in relativistic motion,
the simple
rule that the size is limited to $\sim c \Delta T$ does not apply.
The high $\Gamma$ factor  implies that visible shells are moving
directly towards the observer: if the material of the shell is a
narrow cone, it is unlikely that the observer would be within the
radiation beam yet outside the cone of material (see FMN).

Surprisingly, the curvature of the shell within
$\Gamma^{-1}$ is just as important in determining the envelope of
emission as the overall expansion. This is understood by
distinguishing  the
arrival time of the photons at the detector 
from  the detector's rest frame
time. We denote the former as $T$, and the latter as $t$.
Assume  the shell expands at
velocity $v$ and emits for time $t$.
Because the emitting surface keeps up with the emitted photons, the
photons will arrive at the detector within time
$T = (c-v)t/c \approx  t/(2\Gamma^2)$.
In contrast, the curvature of the shell causes photons emitted from
the material at angle $\theta = \Gamma^{-1}$ to arrive after the photons 
emitted on axis by
$T = v t(1-\cos\theta) \approx t/(2\Gamma^2)$.
Thus, both the overall expansion (which might last $10^7$ sec) and the
delays caused by the curvature spread the observed signal over arrival
times by about $t/(2\Gamma^2)$.
Envelopes should, therefore,  be estimated under the assumption of ``local
spherical symmetry'': local because only $\theta \sim \Gamma^{-1}$ can
contribute, symmetric because the material is seen head on, and
spherical because curvature effects are important.

One can calculate the expected envelope of emission from an expanding
shell. Let $P(\theta,\phi,R)$ give the rate of gamma-ray production
for the shell as a function of spherical coordinates.
  Motivated by the ``external shock''
models~\cite{mr}, we assume a single shell, $R = vt$, which
expands for a time ($t_0$) in a photon quiet phase and then emits from
$t_0$ to $t_{\rm max}$ (i.e., $P(\theta,\phi,R) = P_0$ from $R = vt_0$
to $R = vt_{\rm max}$, and zero otherwise).  In terms of arrival time, the
{\it on-axis} emission will arrive between  $T_0 = t_0/(2\Gamma^2)$ and
$T_{\rm max} = t_{\rm max}/(2\Gamma^2)$.  However, because the
curvature is important, off-axis photons will be delayed,
and  most emission will arrive much later. The
expected envelope, $V(T)$, is:
$$
V(T) = 0  ~~~~~~~~~~~~~~~~~~~~~~~~~~~~~~~~~~~~~~~~~~~~~~~{\rm if}~~T < T_{0}
\eqno(\neweq a)
$$
$$
{}~~~~~ =
{K P_0}
{T^{\alpha+3} - T_{0}^{\alpha+3} \over T^{\alpha+1}}
{}~~~~~~~~~~~~~~~~{\rm if}~T_{0} < T < T_{\rm max}
\eqno(\lasteq b)
$$
$$
{}~ =
{K P_0}
{T_{\rm max}^{\alpha+3} - T_{0}^{\alpha+3} \over  T^{\alpha+1}}
{}~~~~~~~~~~~~~~~~~~~~~~{\rm if}~T > T_{\rm max}
\eqno(\lasteq c)
$$
where $\alpha$ is a typical number spectral index ($\sim 1.5$) and $K$
is a constant.

This envelope is similar to a ``FRED'' (fast rise, exponential decay)
where the fast rise depends mostly on the duration of the photon
active phase
($T_{\rm max}-T_0$) and the slow, power law decay depends mostly on
the duration of the photon quiet phase.  The decay phase is due to
photons delayed by the curvature.  

Often, GRBs do not have a FRED-like shape, implying that   something 
must break the
local spherical symmetry. Perhaps $P(\theta,\phi,R)$ is patchy on
angular scales smaller than $\Gamma^{-1}$, with
each patch contributing an observed peak.  If so, we define the
``filling factor'', $f$, to be the ratio of the observed emission to
what we would expect under local spherical symmetry (see Eq. 32 in FMN):
$$
f = { \int P(\theta,\phi,t)(1-\beta\cos\theta)^{-3} dA  \over
      \int (1-\beta\cos\theta)^{-3} dA }  \eqno(\neweq)
$$
Thus, we propose the ``shell symmetry'' problem for cosmological
GRBs: models incorporating a single release of energy that forms a
relativistic shell must somehow explain  either how the material is
confined to pencil beams narrower than $\Gamma^{-1}$ or how a shell
can have a low filling factor with a correspondingly higher energy
requirement.  

 {}From Eq. 1, we find that the half-width of a GRB, $\sim T_{\rm dur}/2$, is
$\sim T_0/5$.  Thus the shell expands to about $R \sim 5\Gamma^2 T_{\rm
dur}$ before becoming active. In previous work~\cite{mr}, the
photon quiet phase was estimated from 
$E_0 = (\Omega/4\pi)R_{\rm dec}^3 \rho_{\rm ISM}(m_pc^2)\Gamma^2$ where $E_0$
is the energy required to sweep up the ISM with
density $\rho_{\rm ISM}$, $m_p$
is the mass of a proton,
$\Omega$ is the total angular size of the
shell, and   $R_{\rm dec}$ is the radius of the photon quiet phase
where  the shell
decelerates and begins to convert its energy to gamma-rays.
(Note that one cannot solve $E_0$ for $R_{\rm dec}$ with an assumed
$\Gamma$ because $R$ is related to $\Gamma$ through the curvature effects.)
Using  $R = 5\Gamma^2 T_{\rm dur}$,
we find that $E_0$
is an extremely strong function of $\Gamma$:
$
E_0 \sim 10^{32} \Gamma^8 T_{\rm dur}^3 \Omega\rho_{\rm ISM}
$
erg.
Unless $E_0$ is much larger than
$10^{53}$ erg,
$\Gamma$ is quite small ($\sim 75$) for bursts with $T_{\rm dur} \sim
100$ s.

Piran~\cite{piran} has suggested that the filling factor is $\sim 1/N$,
where $N$ is the number of peaks in a burst, and that this filling
factor is so small
that it rules out single relativistic shells in favor of central
engines.
However, it is possible to create many peaks and have a large filling
factor (as in Eq. 2) by allowing for variations in $P(\theta,\phi,R)$
(work in progress).
Thus, we believe it
is too premature to ``rule out'' single relativistic shells. 
Also, there are other ways to overcome inefficiencies. For example,
$\Omega$ might be small.

Shaviv~\cite{shaviv} has suggested that a single shell
sweeps over a cluster of
stars with each star contributing a peak to the time history.  However,
in such a scenario, $T_0$ is effectively zero so the envelope should
have  a rise that scales as $T^2$ (cf. Eq. 1), which is not often seen.
In addition, the Shaviv model
requires $\Gamma \sim 10^3$, so the energy to sweep up the ISM is
extremely large: $10^{62}\rho_{\rm ISM}\Omega$. Globular clusters
will have small $\rho_{\rm ISM}$, but not small enough.
Other issues related to the time history and
emission process  have been raised by Dermer~\cite{dermer}.

We conclude that GRBs do not show the signature of a single
relativistic shell, and models must, therefore, explain how local spherical
symmetry is broken enough to produce the chaotic time histories.

{\it Note added for astro-ph:} In Fenimore, Madras, \& Nayakshin,
equation (1)  was incorrectly dervied
and that was repeated in the manuscript of this paper submitted to the
proceedings.  The error was corrected in
Fenimore \& Sumner (Proc of All-Sky X-ray Observations in the
Next Decade Workshop) and here.
The difference in the equations does not affect our conclusions.

\end{document}